# Grain rotation and lattice deformation during photoinduced chemical reactions revealed by in-situ X-ray nanodiffraction

Zhifeng Huang[1§] Matthias Bartels[2], Rui Xu[1], Markus Osterhoff[2], Sebastian Kalbfleisch[2], Michael Sprung[3], Akihiro Suzuki[5], Yukio Takahashi[5], Thomas N. Blanton[4†], Tim Salditt[2] and Jianwei Miao[1]

[1]Department of Physics & Astronomy and California NanoSystems Institute, University of California, Los Angeles, CA 90095, USA. [2]Institut für Röntgenphysik, Georg-August-Universität Göttingen, Friedrich-Hund-Platz 1, 37077 Göttingen, Germany. [3]DESY, Notkestr. 85, 22607 Hamburg, Germany. [4]Kodak Technology Center, Eastman Kodak Company, Rochester, NY 14650-2106, USA. [5]Department of Precision Science and Technology, Graduate School of Engineering, Osaka University, 2-1 Yamada-oka, Suita, Osaka 565-0871, Japan. [§]Present affiliation: Carl ZEISS X-ray Microscopy Inc., Pleasanton, CA 94588, USA. [†]Present affiliation: International Centre for Diffraction Data, Newtown Square, PA 19073, USA.

*In-situ* **X-ray diffraction (XRD) and transmission electron microscopy (TEM) have been used to investigate many physical science phenomena, ranging from phase transitions, chemical reaction and crystal growth to grain boundary dynamics[1-7]. A major limitation of** *in-situ* **XRD and TEM is a compromise that has to be made between spatial and temporal resolution[1-7]. Here, we report the development of** *in-situ* **X-ray nanodiffraction to measure atomic-resolution diffraction patterns from single grains with up to 5 millisecond temporal resolution, and make the first real-time observation of grain rotation and lattice deformation during photoinduced chemical reactions: $Br^- + h\nu \rightarrow Br + e^-$ and $e^- + Ag^+ \rightarrow Ag^0$. The grain rotation and lattice deformation associated with the chemical reactions are quantified to be as fast as 3.25 rad./sec. and as large as 0.5 Ångstroms,** **respectively. The ability to measure atomic-resolution diffraction patterns from individual grains with several millisecond temporal resolution is expected to find broad applications in materials science, physics, chemistry, and nanoscience.**

Many materials are polycrystalline and are made of a large number of grains of varying size and orientation. The structure and dynamics of grains and grain boundaries are thus fundamental to many materials properties[8,9]. Several experimental methods can be used to characterize grains and grain boundaries in materials, including X-ray diffraction (XRD), transmission electron microscopy (TEM), electron tomography, scanning electron microscopy (SEM), electron diffraction, and optical microscopy[4,5,8-15]. Although TEM and electron tomography can image grains and grain boundaries at atomic scale resolution[10,14,15], the temporal resolution in probing individual grain dynamics with *in situ* XRD or TEM remains very limited[4,5,11,13,16]. This limitation stems from the fact that, in order to investigate the fast dynamics of individual grains, an intense beam must be focused onto single grains in materials and the diffraction patterns have to be collected by an area detector with a high dynamic range, high quantum efficiency and fast readout time. Here, we demonstrate *in situ* X-ray nanodiffraction to measure atomic resolution diffraction patterns from single grains with several millisecond temporal resolution through a combination of a brilliant synchrotron undulator beam, Kirkpatrick-Baez (K-B) mirrors and a state-of-the-art X-ray detector (PILATUS). With this system, we observe, for the first time, grain rotation and lattice deformation during photoinduced chemical reactions with up to 5ms temporal resolution.

Figure 1a shows the schematic layout of the *in situ* X-ray nanodiffraction instrument, utilizing presently the brightest synchrotron in the world – the Positron-Electron Tandem Ring Accelerator (PETRA) III in Germany. Monochromatic X-rays with E = 13.8 or 13.6 keV were focused to a spot of ~370nm×270nm by two K-B mirrors[17]. The sample is positioned at the focal spot with a flux of ~6.74×10$^{11}$ photons/sec. Two detector configurations are implemented in the experiments. In configuration one, a PILATUS 1M detector, consisting of 10 modules each with 487×195 pixels and a pixel size of 172×172 μm$^2$, is placed at a distance of 167.8mm downstream of the sample. The PILATUS detector has a dynamic range of 20 bits and single-photon sensitivity[18]. The readout time of PILATUS 1M is 40ms in the full-frame mode, 9ms in the 3-module mode, and 3ms in the single-module mode. In configuration two, a PILATUS 6M detector of 60 modules is mounted at a distance of 386.7mm from the sample. The readout time of PILATUS 6M is 30ms in the full-frame mode. This instrument allows us to measure atomic resolution diffraction patterns with millisecond time scales from a ~370nm×270nm illumination area (Methods and Supplementary Fig. 1).

Using this *in situ* X-ray nanodiffraction system, we have studied three groups of samples in an ambient environment at room temperature (Methods): i) Kodak direct print linagraph paper (Type 2167) containing AgBr, gelatin and other materials; ii) AgBr control samples consisting of either AgBr powder on $Si_3N_4$ membranes or AgBr/gelatin; and iii) non-AgBr control samples consisting of $TiO_2$/gelatin, $CeO_2$/gelatin or Ag/gelatin. Figure 1b and Supplementary Video 1 show real-time measurements of the diffraction patterns from a Kodak linagraph paper during photoinduced chemical reactions ($Br^- + h\nu \rightarrow Br + e^-$ and $e^- + Ag^+ \rightarrow Ag^0$), initiated by exposing the sample to X-rays. In the Kodak paper, cubic AgBr grains with an average size of ~700nm are distributed in a layer near the surface (Supplementary Fig. 2). Because the size of the X-ray beam is smaller than the individual AgBr grain size, several large and high intensity diffraction spots are initially diffracted from single AgBr grains. With more X-ray irradiation, the large AgBr (200) and (220) spots gradually become smaller and increase in number of diffraction spots, indicating the AgBr grains are decomposed into smaller grains. Meanwhile, Ag (111) and (200) diffraction spots appear in the diffraction patterns, implying that Ag nano-grains start nucleating and growing with the photolysis of the AgBr grains. After several seconds of X-ray exposure, the Debye-Scherrer rings of AgBr (200), (220), Ag (111) and (200) start to appear (Fig. 1b, Supplementary Fig. 3 and Video 1). These represent real-time measurements of the photolysis of AgBr to produce Ag[19,20]. This type of the photolysis is also observed by exposing controlled AgBr samples to the nano-focused X-rays (Supplementary Fig. 4 and Video 2). Compared to the controlled samples, the speed of the photolysis of AgBr in the Kodak paper is faster, which is due to the chemical sensitization of photographic Ag halides grains (Supplementary Information), a more uniform distribution and smaller size of AgBr grains in the Kodak paper. For the controlled non-AgBr samples, we do not observe any motion of the diffraction spots.

Next, we investigate the occurrence of grain rotation during the chemical reaction of AgBr with photons to produce Ag. As the PILATUS detector captures a cross section of the Ewald sphere (i.e. forming a Debye-Scherrer ring), evidence of the grain rotation is observed in both the Kodak linagraph paper and controlled AgBr samples (Supplementary Videos 1 and 2). To quantitatively analyze the grain rotation, we azimuthally plotted the intensity of the AgBr (200) and Ag (111) Debye-Scherrer rings as a function of the exposure time for Supplementary Video 1 (Fig. 2a,b). Careful examination of the figures indicates that there are three types of grain-related features: 1) Point-like tracks, representing a momentary appearance and disappearance of the diffraction spots, i.e., grain rotation tracks intersecting the detector plane; 2) horizontal-line-like tracks, representing stationary diffraction spots and grains; 3) Curve-line-like tracks along the Debye-Scherrer ring, representing grain rotation around the axis perpendicular to the detector plane. In particular, some of the curve-line-like tracks are oblique lines, indicating a constant angular velocity of the grain rotation.

While grain rotation due to external load or during grain growth has been previously described[4,11,16,21,22], in this study we observed the grain rotation during chemical reactions that has not been previously reported. According to photographic theory[19,20] (Supplementary Information), defects (i.e. points and dislocations) in AgBr grains act as sensitized sites for Ag nucleation and aggregation. Unlike visible light which produces one Ag atom per absorbed light photon, an absorbed X-ray photon can produce many photolytic Ag atoms[19], which can grow and agglomerate on the surfaces of and inside a AgBr grain. As latent imaging sites grow in size due to Ag atom agglomeration in different locations inside the AgBr grain, stress is generated inhomogeneously. Once the resulting strain exceeds elastic deformation, the AgBr grain can fragment into smaller AgBr grains, which accompanies with the growth of Ag nano-grains. Figure 2c shows a major AgBr (200) spot decomposed into smaller ones. SEM images of both the Kodak linagraph paper and controlled AgBr samples confirm the fragmentation of AgBr grains and the existence of filamentary Ag structure on the surface of the grains after X-ray exposure (Supplementary Fig. 5), which is also consistent with post light exposure studies of Ag halides[23,24]. With the release of the internal stress, the angular velocity of the major diffraction spot gradually slows down as a function of the exposure time (Fig. 2d) and eventually becomes stable (Fig. 2a). In the experiments of this study, the angular velocity of the grain rotation is measured to be as fast as 3.25 rad./sec (Fig. 2e, Supplementary Video 3), which is significantly faster than that of previously reported grain rotation[4,11,16,21,22].

In addition to grain rotation, lattice deformation is also observed during the X-ray exposure of the Kodak linagraph paper and controlled AgBr samples. Real-time measurements of simultaneous grain rotation and lattice deformation in photolysis of AgBr to produce Ag grains are shown in Fig. 3, Supplementary Figs. 6 and 7 and Videos 4-8. In our experiments, two types of Ag lattice deformation are observed. The first is irregular lattice deformation, i.e. diffraction spots are irregularly oscillated around the Ag Debye-Scherrer rings. Figure 3b and Supplementary Video 4 show a diffraction spot initially observed with a lattice spacing of 2.363 Å, which is close to the Ag (111) spacing of 2.361 Å. The spot then quickly moves to lower $2\theta$ direction (i.e. lattice expansion). After the lattice spacing reaches a maximum of 2.397 Å, the diffraction spot reverses its direction and moves to higher $2\theta$ direction (i.e. lattice contraction). After 1.36 sec. the lattice spacing reaches a minimum of 2.309 Å and then the diffraction spot moves back to the Ag (111) ring again (i.e. lattice expansion) until the spot finally disappears on the detector plane. The time span of the whole lattice deformation process is 2.29 sec., which is simultaneously accompanied with grain rotation. Figure 3e shows the lattice spacing, the corresponding normal component strain, and the angular distribution of the diffraction spot as a function of the exposure time. The normal component strain is determined to be in the range of −2.22% to +1.51% in this example.

The second type of lattice deformation is more regular, i.e. the tracks of diffraction spots form straight or curved lines across the Ag (111) and (200) Debye-Scherrer rings. Figure 3d and Supplementary Video 6 show a diffraction spot initiating on the Ag (200) ring and moving in a straight line towards inside the Ag (111) ring, which corresponds to a lattice expansion of 0.424 Å. The diffraction spot then moves from inside the Ag (111) ring towards outside the Ag (200) ring, corresponding to a lattice contraction of 0.5 Å (Fig. 3f). The whole process of the photoinduced lattice deformation lasts 8.45 sec. and the magnitude of the lattice deformation is significantly larger than those previously reported[25-28]. Like the first track type (Fig. 3e), the second track type of lattice deformation is also simultaneously accompanied with grain rotation (Fig. 3f). Similar tracks are also shown in Supplementary Figs. 6, 7 and Videos 5, 7 and 8. As Ag atoms aggregate to form nanoparticles at the latent imaging sites inside a AgBr grain, stress is accumulated. When the stress exceeds the elastic deformation of the AgBr grain, Ag nano-grains are squeezed out to the surface of the AgBr grain to form filamentary structures[29] (Supplementary Fig. 5). The lattice structure of newly formed Ag nano-grains is generally unstable inside the AgBr grain[30], and the subsequent inhomogeneous large strain developed during the course of being squeezed out to the surface of the AgBr grain introduces lattice contraction or expansion.

In conclusion, we demonstrate an *in situ* X-ray nanodiffraction system with up to 5ms temporal resolution, and perform real-time measurements of the grain rotation and lattice deformation during photoinduced chemical reactions. As both advanced synchrotron light sources and X-ray optics are currently under rapid development worldwide[31], the focal spot of the *in situ* X-ray nanodiffraction system can be further improved to the ~10 nm level, allowing the study of the sub-grain dynamics in materials. We thus anticipate that such an *in situ* X-ray nanodiffraction technique will find broad applications across several disciplines.

**Methods**

**Sample preparation**

*Group #1: Kodak direct print linagraph paper:*

Kodak direct print linagraph paper (Type 2167) turns to light gray after exposure to visible light, and turns to black after exposure to X-rays. The Kodak linagraph paper uses cellulose Type Ib as the support and gelatin as the coating binder, and contains talc (paper transport aid), hydrotalcite (paper transport aid), rutile $TiO_2$ (filler), polyethylene (overcoat) and AgBr (photosensitive component). An XRD phase analysis of the linagraph paper measured with a Cu rotating anode X-ray generator is shown in Supplementary Fig. 3b.

*Group #2: AgBr control samples*

All AgBr control samples were made under low light conditions to reduce the amount of Ag print out.

1) AgBr/membrane

Into a glass beaker was added 50 ml deionized (DI) water and 3.167g $AgNO_3$ (Eastman Kodak Company). Into a second beaker was added 50 ml DI water and 2.124g NaBr (MCB Chemical). The $AgNO_3$ solution was added drop wise to the NaBr solution while stirring with a magnetic stir bar. A precipitate immediately formed. The resulting dispersion was allowed to stir 15 min and was then vacuum filtered using a Millipore RA filter (1.2 μm). The collected solids were transferred to a glass dish and dried 30 min at 100°C. XRD identified the dried powder as AgBr. The AgBr powder was mixed with distilled water, and a droplet of solution was placed on a silicon nitride membrane. The membrane was air dried before the *in situ* XRD experiment.

2) AgBr/gelatin

Into a glass beaker was added 75 ml DI water and 0.45g $AgNO_3$. Into a second beaker was added 50 ml DI water and 0.29g NaBr. The $AgNO_3$ solution was added drop wise to the NaBr solution while stirring with a magnetic stir bar. A precipitate immediately formed. The resulting dispersion was allowed to stir 15 min and was then centrifuged. The liquid centrifugate was poured off, DI water added, sonicated and the centrifuge process was repeated. To the remaining solids in the centrifuge tube was added 10 ml of a 5 weight percent Type 40 gelatin (Eastman Gelatin) aqueous solution (gelatin solution at 40°C). The solution was sonicated and then poured into a glass beaker, while kept at 40°C and stirred using a magnetic stir bar. Horiba analysis found the median AgBr particle size to be ~0.134 μm and an XRD experiment confirmed the particles to be AgBr. AgBr/gelatin samples for this study were prepared as free standing films or on a filter. An aliquot of the AgBr – gelatin dispersion was puddle deposited onto an unsubbed poly(ethylene terephthalate) (PET) support. A draw bar with a 200 micron gap was used to produce a wet coating of the dispersion on the PET. The coating dried in ambient air, and was removed from the PET resulting in a free-standing AgBr – gelatin film. A 4 ml aliquot of the AgBr – gelatin dispersion from above was vacuum filtered using a Millipore VM filter (0.05 micron). The collected sample was washed with 100 ml of 40 °C DI water to remove most of the gelatin. The filter with the AgBr coating was removed from the filter apparatus and allowed to dry in ambient air.

*Group #3: non-AgBr control samples*

These samples were used to confirm that the grain rotation and lattice deformation observed are not due to gelatin.

1) $TiO_2$/gelatin

   Into a glass beaker was added 99.91g of a 5 weight percent Type 40 gelatin aqueous solution (gelatin solution at 40°C) and 0.510g $TiO_2$ (Unitane). The resulting dispersion was allowed to stir 15 min. An aliquot of the $TiO_2$ – gelatin dispersion was puddle deposited onto an unsubbed poly(ethylene terephthalate) (PET) support. A draw bar with a 200 μm gap was used to produce a wet coating of the dispersion on the PET. The coating dried in ambient air, and was removed from the PET resulting in a free-standing $TiO_2$ – gelatin film. Horiba analysis found the median $TiO_2$ particle size to be ~0.480 μm and an XRD experiment confirmed the particles were rutile $TiO_2$.

2) $CeO_2$/gelatin

   Into a glass beaker was added 99.87g of a 5 weight percent Type 40 gelatin aqueous solution (gelatin solution at 40°C) and 0.508g $CeO_2$ (Aldrich nano <25nm). The resulting dispersion was allowed to stir 15 min. An aliquot of the $CeO_2$ – gelatin dispersion was puddle deposited onto an unsubbed poly(ethylene terephthalate) (PET) support. A draw bar with a 200 μm gap was used to produce a wet coating of the dispersion on the PET. The coating dried in ambient air, and was removed from the PET resulting in a free-standing $CeO_2$ – gelatin film. An XRD experiment confirmed the particles were cerianite $CeO_2$.

3) Ag/gelatin

   Into a glass beaker was added 99.74g of a 5 weight percent Type 40 gelatin aqueous solution (gelatin solution at 40°C) and 0.501g Ag metal powder (Aesar). The resulting dispersion was allowed to stir 15 min. An aliquot of the Ag – gelatin dispersion was puddle deposited onto an unsubbed poly(ethylene terephthalate) (PET) support. A draw bar with a 200 μm gap was used to produce a wet coating of the dispersion on the PET. The coating dried in ambient air, and was removed from the PET resulting in a free-standing Ag – gelatin film. Horiba analysis found the median Ag particle size to be ~24.8 μm and an XRD experiment confirmed the particles were Ag.

**Experimental set-up.** At the Coherence Beamline P10 at the PETRA III synchrotron, the source of X-rays in low beta configuration is a 5 m long U29 undulator. The X-rays are monochromatized by a Si (111) double crystal monochromator at 35 m away from the source. The monochromatic X-ray beam is focused by a pair of K-B mirrors of elliptical shape with a Pd coating. The horizontally focusing mirror (HFM, by JTEC) has a focal length of 200 mm and height errors of 4 nm (peak to valley); the vertically focusing mirror (VFM, by WinlightX) has a focal length of 305 mm, with height errors around 10 nm. The X-ray beam hits the mirrors under a grazing angle of incidence of 4 mrad at the center; the active length is about 80 mm. At the exit window of the KB mirror vacuum tank, a coarse W aperture is used to filter only the double-reflected parts of the beam, and suppresses X-rays that hit only one or no mirror at all.

Two tantalum apertures are placed at the upstream of the focus to reduce the scattering from the K-B mirrors The focus is measured and tweaked by scanning a crossed pair of one dimensional X-ray waveguides[32], by recording the transmitted intensity with a silicon PIN diode (Canberra PD300-500CB), with a thickness of 500 µm and an active area of 19 mm in diameter. Each pair of waveguides consists of a C guiding layer of 35 nm thick, a Mo interlayer and a Ge cladding material; the waveguides are cut to a length of some hundred micrometers. Absorbed X-ray photons will generate a number of electron-hole pairs of 3.66 eV. The resulting charge separation is measured as a current with a picoammeter (Keithley 6485). Two PILATUS detectors are used in two independent experimental configurations. A PILATUS 1M detector with 981×1043 pixel active area and totally 10 modules is placed at a distance of 167.8 mm downstream of the sample, detecting 13.8 KeV monochromatic X-rays. In the second configuration, a 6M detector with 2463×2527 pixel active area and a total of 60 modules is placed at a distance of 386.7 mm from the sample, detecting 13.6 KeV monochromatic X-rays. The distance between the sample and the detector was calibrated by the diffraction patterns of a Ag foil (Supplementary Fig. 1).

**Acknowledgments:** We thank I. Vartaniants for stimulating discussions. This work was supported by the DARPA PULSE program through a grant from AMRDEC, UC Discovery/TomoSoft Technologies (IT107-10166), Helmholtz Society grant VH-VI-403 and DFG SFB755.

**Author information** The authors declare no competing financial interests. Correspondence and requests for materials should be addressed to J. M. (miao@physics.ucla.edu).

**Figure legends**

**Figure 1**. **Schematic layout of an *in situ* X-ray nanodiffraction system with up to 5 millisecond temporal resolution. a**, Monochromatic X-rays are focused to a spot of ~370nm×270nm (inset) by two K-B mirrors. A pinhole is used to remove the parasitic scattering from the K-B mirrors. The sample is positioned at the focal spot with a flux of ~$6.74 \times 10^{11}$ photons/sec, and atomic resolution diffraction patterns from the samples are collected by a PILATUS 1M or 6M detector. Three representative diffraction patterns are measured from a Kodak linagraph paper at 0, 0.84 and 2.66 sec., respectively, in which diffraction spots (black arrows) are rotating along the Debye-Scherrer rings. **b**, Real-time observation of the photolysis of AgBr grains to produce Ag nano-grains. Four representative diffraction patterns excerpted from Supplementary Video 1 at 0, 0.98, 3.50 and 6.16 sec., respectively, which are collected by the PILATUS 1M detector (temporal resolution: 140ms).

**Figure 2. Real-time observation of grain rotation during photoinduced chemical reactions. a**,**b**, Azimuthal plot of the diffraction intensity of the AgBr (200) and Ag (111) Debye-Scherrer rings as a function of the exposure time for Supplementary Video 1 (temporal resolution: 140ms). Three types of grain related features are visible: 1) Point-like tracks; 2) Horizontal-line-like tracks; and 3) Curve-line-like tracks along the Debye-Scherrer ring (arrows). **c**, A sequence of images of the rectangular region in (**a**), showing a major AgBr (200) spot fragments into smaller ones. **d**, Measurements of the angular velocity of a major diffraction spot in (**c**) as a function of the time. **e**, Measurements of the angular velocity of a diffraction spot in Supplementary Video 3 (temporal resolution: 5ms), in which the fastest instantaneous angular velocity is 3.25 rad./sec.

**Figure 3. Real-time observation of simultaneous grain rotation and lattice deformation during photoinduced chemical reactions. a**, Diffraction intensity averaged from real-time measurements of a Kodak linagraph paper (Supplementary Videos 4 and 5, temporal resolution: 29ms). **b**, A track image and 7 representative frames of the rectangular region in (**a**) extracted from Supplementary video 4. The track image shows the trajectory of a moving diffraction spot between 10.50 and 12.79 sec.,

where arrows show the motion of the diffraction spot. **c**, Diffraction intensity averaged from real-time measurements of a controlled AgBr/membrane sample (Supplementary video 6, temporal resolution: 130ms). **d**, A track image and 18 representative frames of the rectangular region in (**c**) extracted from Supplementary Video 6. The track image shows the trajectory of a diffraction spot between 31.33 and 39.78 sec., where arrows show the motion of the diffraction spot. **e**, Quantification of the lattice deformation, normal component strain and grain rotation of the rectangular region in (**a**) between 10.50 and 12.79 sec., where labels (a)-(g) correspond to those in (**b**). **f**, Quantification of the lattice deformation and grain rotation of the rectangular region in (**c**) between 31.33 and 39.78 sec., where labels (a)-(q) correspond to those in (**d**).

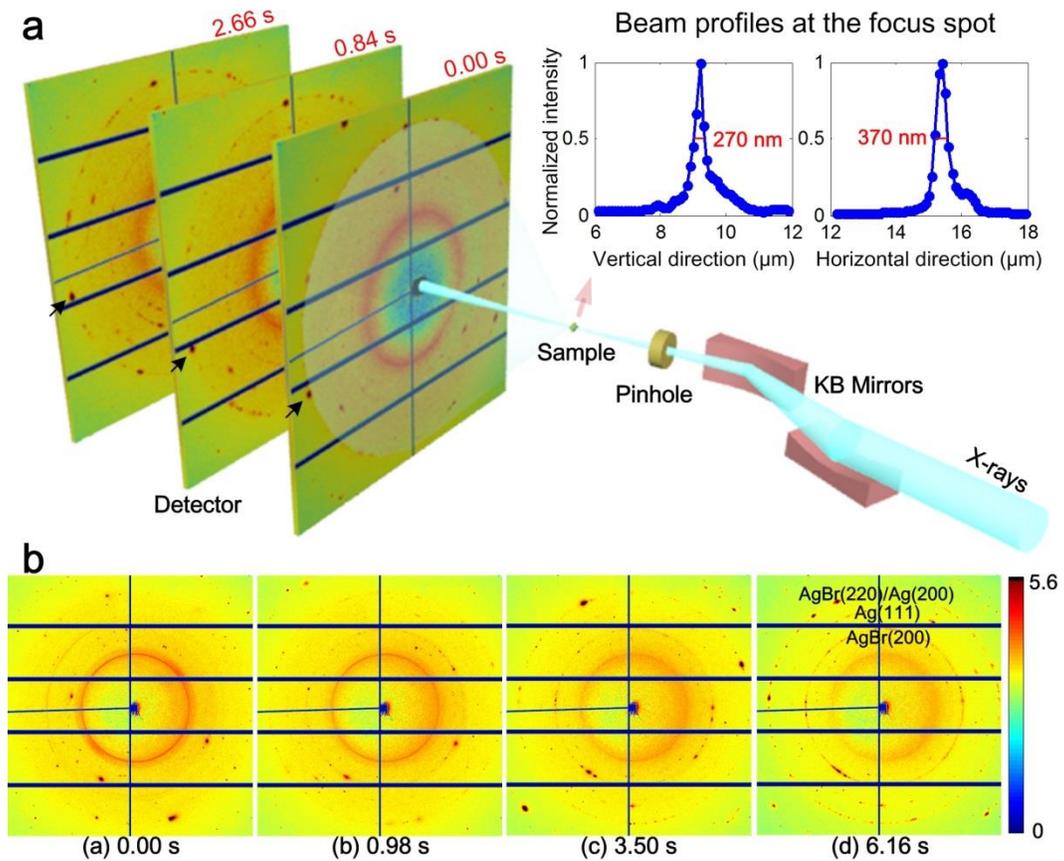

**FIG. 1**

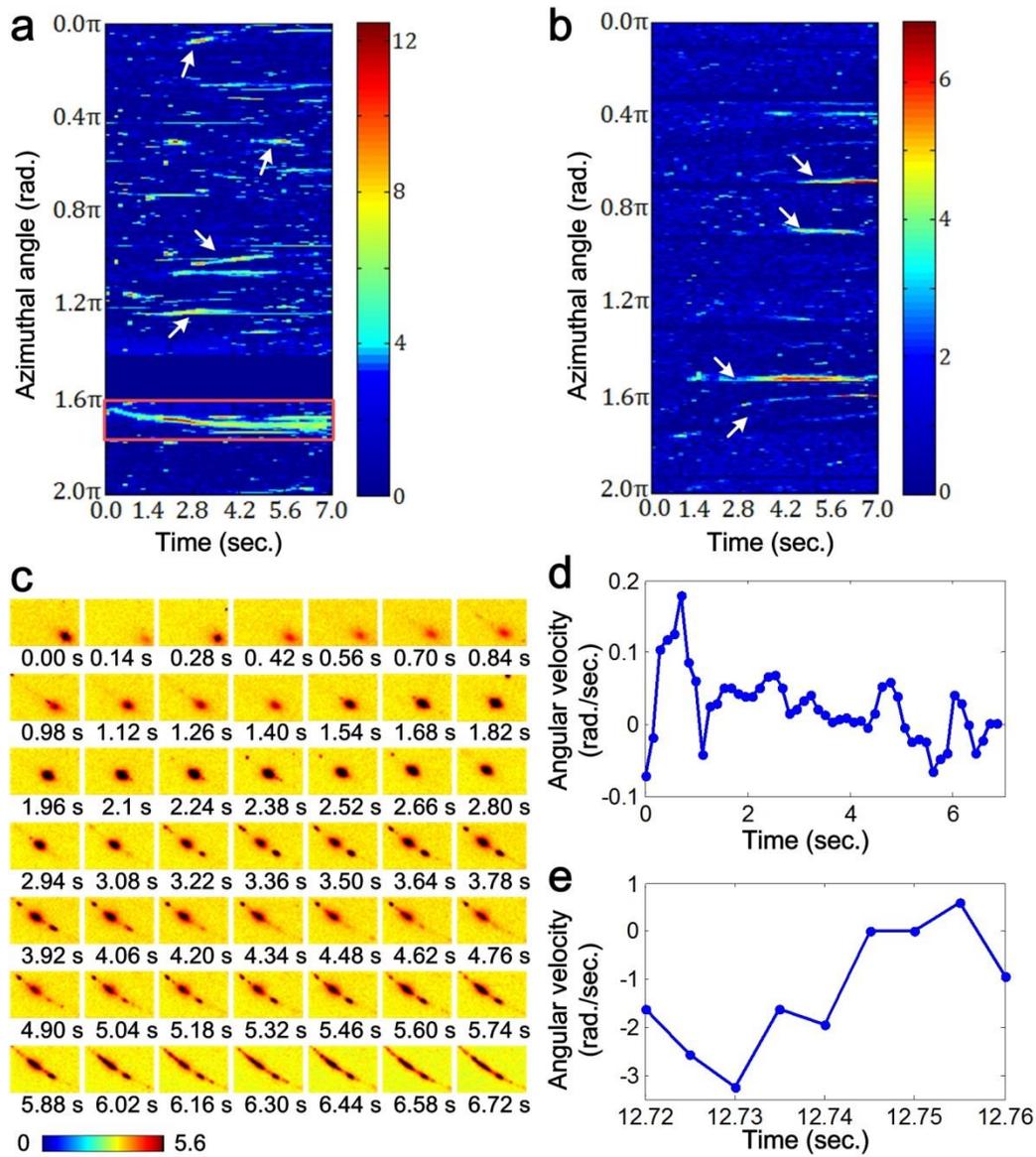

**FIG. 2**

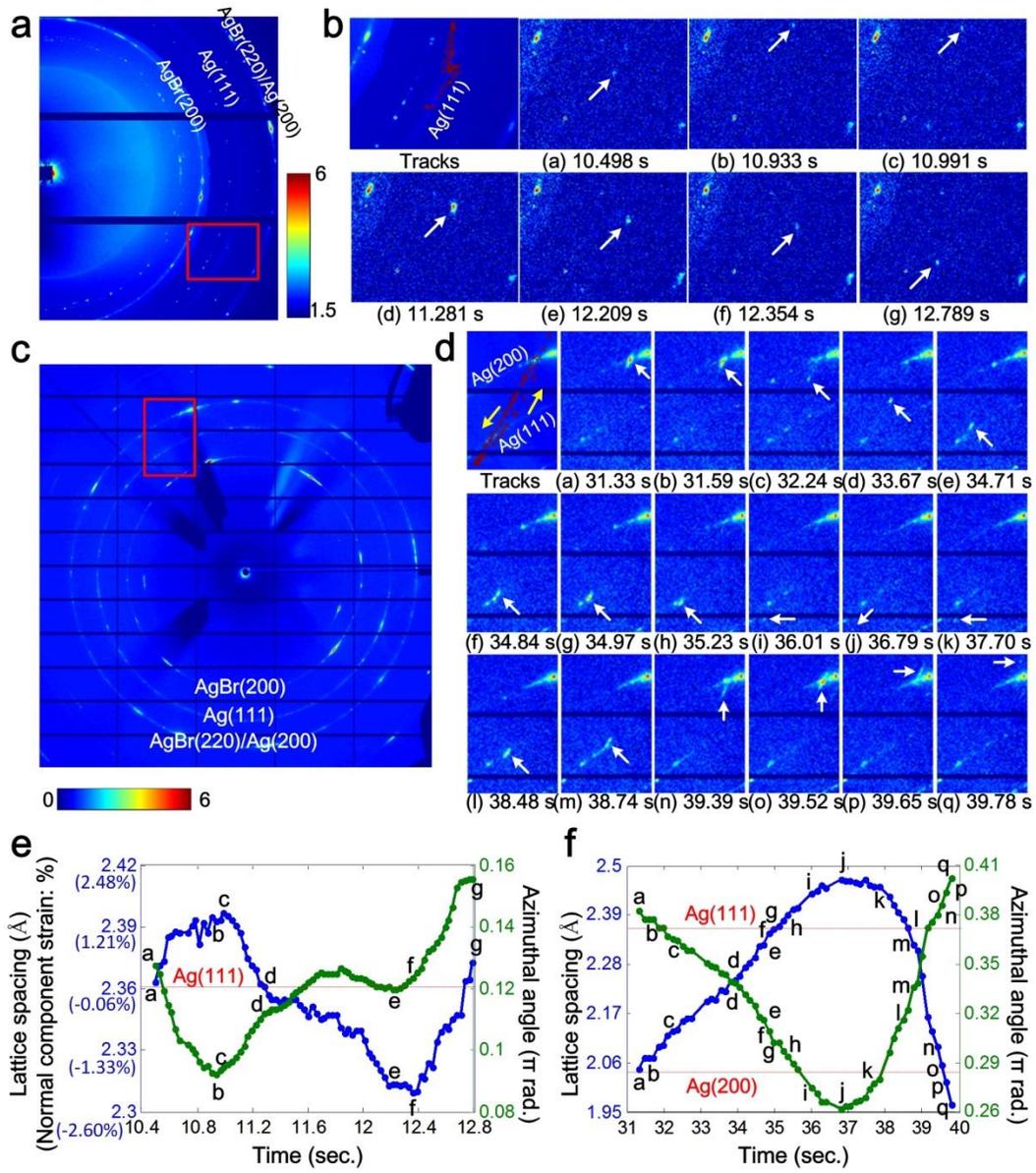

**FIG. 3**